\documentclass{aastex63}

\usepackage{tikz}
\usetikzlibrary{arrows}
\usepackage{amsmath}
\usepackage{comment}

\shorttitle{Thermal Emission in finite atmosphere}

\graphicspath{{./}{figures/}}

\begin{document}

\title{Atmospheric Thermal Emission Effect on Chandrasekhar's Finite Atmosphere Problem}

\correspondingauthor{Soumya Sengupta}
\email{soumya.s@iiap.res.in}

\author[0000-0002-7006-9439]{Soumya Sengupta}
\affiliation{Indian Institute of Astrophysics, Koramangala 2nd Block,
Sarjapura Road, Bangalore 560034, India}
\affiliation{Pondicherry University, R.V. Nagar, Kalapet, 605014, Puducherry, India}

\begin{abstract}
    The solutions of the \textit{diffuse reflection finite atmosphere problem} are very useful in the astrophysical context. Chandrasekhar was the first to solve this problem analytically, by considering atmospheric scattering. These results have wide applications in the modeling of planetary atmospheres. However, they cannot be used to model an atmosphere with emission. We solved this problem by including \textit{thermal emission effect} along with scattering. Here, our aim is to provide a complete picture of generalized finite atmosphere problem in  presence of scattering and thermal emission, and to give a physical account of the same. For that, we take an analytical approach using the invariance principle method to solve the diffuse reflection finite atmosphere  problem in the presence of atmospheric thermal emission. We established the general integral equations of modified scattering function $S(\tau; \mu, \phi; \mu_0, \phi_0)$, transmission function $T(\tau; \mu, \phi; \mu_0, \phi_0)$ and their derivatives with respect to $\tau$ for a thermally emitting atmosphere. We customize these equations for the case of isotropic scattering and introduce two new functions $V(\mu)$, and $W(\mu)$, analogous to Chandrasekhar’s $X(\mu)$, and $Y(\mu)$ functions respectively. We also derive a transformation relation between the modified S-T functions and give a physical account of $V(\mu)$ and $W(\mu)$ functions. Our final results are consistent with those of Chandrasekhar at low emission limit (i.e. only scattering). From the consistency of our results, we conclude that the consideration of thermal emission effect in diffuse reflection finite atmosphere  problem gives more general and accurate results than considering only scattering.
\end{abstract}


\section{Introduction}\label{sec: intro}

Chandrasekhar did the pioneering work on the process of radiative transfer which is the heart of the observations as well as modeling in astrophysical context \cite{chandrasekhar1960radiative}. One of his most interesting and useful method is \textit{the Invariance Principle} technique, which has a great deal of applications in atmospheric modeling. Although this principle was first introduced by \cite{ambartsumian1943cr,ambartsumian1944problem}, \cite{chandrasekhar1960radiative} used this theory to solve the semi-infinite and finite atmosphere problem in its most elegant way by introducing the scattering function $S(\tau;\mu,\phi;\mu_0,\phi_0)$ and transmission function $T(\tau;\mu,\phi;\mu_0,\phi_0)$. The final results of those treatments can be represented in terms of H-function (semi-infinite case) \cite{chandrasekhar1947radiative1} and X-, Y- functions (finite case)  \cite{chandrasekhar1948radiative}. The values of the H-function \citep{chandrasekhar1947radiative2} and X-,Y- functions \citep{chandrasekhar1952x1,chandrasekhar1952x2} in the case of isotropic scattering are directly used in atmosphere modeling. Even a simple transformation rule between S and T was established by \cite{coakley1973simple}.

Although the results provided in \cite{chandrasekhar1960radiative} have direct applications in stellar and planetary problems, the treatment is not complete in some sense as it does not consider the atmospheric emission and scattering simultaneously. \cite{bellman1967chandrasekhar} include the thermal emission in the planetary atmosphere problem and started a new technique called invariant embedding \citep{bellman1992introduction}. In the context of exoplanetary transmission spectra modeling, \cite{sengupta2020optical} and \cite{chakrabarty2020effects} showed the crucial effect of scattering and atmospheric re-emission respectively. Recently, \cite{sengupta2021effects} considered scattering and atmospheric emission simultaneously to study the modifications in Chandrasekhar's semi-infinte atmosphere problem. However, the effect of emission on the finite atmosphere problem, which is more general than the semi-infinite one, remains unsolved. 

In this work we solve the finite atmosphere problem in case of isotropic scattering and emission by the same analytical procedure as shown in \cite{sengupta2021effects}. For that we consider local thermodynamic equilibrium condition in vertical atmospheric layers, which ensures the fact that each layer contributes the Blackbody emission according to Kirchoff's law \citep{chandrasekhar1960radiative,seager2010exoplanet}. We used the invariance principle method \citep{ambartsumian1944problem,chandrasekhar1960radiative} to derive the modified scattering, transmission function and final radiation to show that our results are more general than Chandrasekhar's results. This treatment is also free from the isothermal atmosphere condition, which was a limitation of the work in \cite{sengupta2021effects}.

 In section~\ref{sec: Invariance Principle for Finite atmosphere} we state the mathematical formula of invariance principles for finite atmosphere following \cite{chandrasekhar1960radiative}. Section~\ref{sec: general integral equations} is devoted to deriving the general integral equations of the Scattering function (S) and Transmission function (T) in case of thermal emission with scattering. The modified form of these functions specifically for the isotropic scattering case is shown in section~\ref{sec: specific form of integral equations}. Then we establish a simple transformation rule between the $S(\mu)$-$T(\mu)$ in section~\ref{sec: transformation rule} and give their physical interpretations in section~\ref{sec: physical meaning}. The consistency of our new results with the literature is discussed in section~\ref{sec: consistency} and concluded with an elaborated discussion in section~\ref{sec: Discussion}.

\section{Invariance Principle for Finite atmosphere}\label{sec: Invariance Principle for Finite atmosphere}
The radiative transfer equation in case of plane-parallel approximation can be written as,
\begin{equation}\label{eq: radiative transfer}
    \mu \frac{dI_\nu(\tau_\nu,\mu,\phi)}{d\tau_\nu} = I_\nu(\tau_\nu,\mu,\phi) - \xi_\nu(\tau,\mu,\phi)
\end{equation}
Here $I_\nu(\tau_\nu,\mu,\phi)$ is the specific intensity at a particular frequency $\nu$, direction cosine $\mu$, angle of azimuth $\phi$ and optical depth range between $\tau_\nu$ to $\tau_\nu + d\tau_\nu$. With these same parameters the source function is written as, $\xi_\nu(\tau_\nu,\mu,\phi)$.

For an atmosphere with simultaneous scattering and absorption, the optical depth can be defined as \citep{domanus1974fundamental,sengupta2020optical,sengupta2021effects},

\begin{equation}\label{eq: optical depth}
    d\tau_\nu = -[\kappa_\nu(z) + \sigma_\nu(z)]dz = -\chi_\nu(z)dz
\end{equation}
Here $\kappa_\nu(z),\sigma_\nu(z)$ and $\chi_\nu(z)$ are the volumetric absorption co-efficient, scattering co-efficient and extinction co-efficient at a particular frequency $\nu$ and depth z respectively.  

Note that, for the sake of simplicity in further calculations we suppress the subscript $\nu$ by considering all the calculations at a particular frequency. It should not be confused with the Grey atmosphere approximation as there is no such assumption in the present work.

Now  finite atmosphere is bounded by optical depth $\tau = 0$ to $\tau=\tau_1$ \citep{chandrasekhar1960radiative}. To provide a solution of the problem of only diffuse reflection from such an atmosphere, \cite{chandrasekhar1947radiative} used the invariance principle method. We will use the same methodology following \citep{chandrasekhar1960radiative} to get a solution of the more general problem where atmospheric thermal emission is also included with the diffuse scattering.

When a radiation of light $\pi F$ incident on an atmosphere of optical thickness $\tau_1$ along the direction ($-\mu_0,\phi_0$), then the diffusely reflected and transmitted intensities can be represented as,
\begin{equation}\label{eq: diffuse reflection1}
    I(0,\mu,\phi) = \frac{F}{4\mu}S(\tau_1,\mu,\phi;\mu_0,\phi_0);
\hspace{0.5cm}
    I(\tau_1,-\mu,\phi) = \frac{F}{4\mu}T(\tau_1,\mu,\phi;\mu_0,\phi_0)
\end{equation}
respectively, where $S(\tau_1,\mu,\phi;\mu_0,\phi_0)$ and $T(\tau_1,\mu,\phi;\mu_0,\phi_0)$ are the scattering and transmission functions. Note that, these two intensities refer only that light which has suffered at least one scattering process and do not include any direct transmission along $(-\mu_0,\phi_0)$ direction. For a detailed discussion on this, we refer \textit{Radiative Transfer} by \cite{chandrasekhar1960radiative}.

The four mathematical expressions of invariance principle in finite atmosphere problem can be written as \citep{chandrasekhar1960radiative},

\textbf{Principle I}
\begin{equation}\label{eq: Principle I}
I(\tau,+\mu,\phi) = \frac{F}{4\mu}e^{-\tau/\mu_0}S(\tau_1-\tau,\mu,\phi;\mu_0,\phi_0) + \frac{1}{4\pi\mu}\int_0^1\int_0^{2\pi}I(\tau,-\mu',\phi')S(\tau_1-\tau,\mu,\phi;\mu',\phi')d\mu'd\phi'
\end{equation}
\textbf{Principle II}
\begin{equation}\label{eq: Principle II}
I(\tau,-\mu,\phi) = \frac{F}{4\mu}T(\tau,\mu,\phi;\mu_0,\phi_0) + \frac{1}{4\pi\mu}\int_0^1\int_0^{2\pi}I(\tau,+\mu',\phi')S(\tau,\mu,\phi;\mu',\phi')d\mu'd\phi'
\end{equation}
\textbf{Principle III}
\begin{equation}\label{eq: Principle III}
\begin{split}
\frac{F}{4\mu}S(\tau_1;\mu,\phi;\mu_0,\phi_0) = \frac{F}{4\mu}S(\tau;\mu,\phi;\mu_0,\phi_0)+e^{-\tau/\mu}I(\tau,+\mu,\phi) + \frac{1}{4\pi\mu}\int_0^1\int_0^{2\pi}I(\tau,+\mu',\phi')T(\tau,\mu,\phi;\mu',\phi')d\mu'd\phi'
\end{split}
\end{equation}
\textbf{Principle IV}
\begin{equation}\label{eq: Principle IV}
\begin{split}
\frac{F}{4\mu}T(\tau_1;\mu,\phi;\mu_0,\phi_0) =& \frac{F}{4\mu}e^{-\tau/\mu_0}T(\tau_1-\tau;\mu,\phi;\mu_0,\phi_0)+e^{-(\tau_1-\tau)/\mu}I(\tau,-\mu,\phi)\\ &+ \frac{1}{4\pi\mu}\int_0^1\int_0^{2\pi}I(\tau,-\mu',\phi')T(\tau_1-\tau,\mu,\phi;\mu',\phi')d\mu'd\phi'
\end{split}
\end{equation}
These equations are derived and diagramatically shown in \cite{chandrasekhar1947radiative,chandrasekhar1960radiative,peraiah2002an}. The boundary conditions used to calculate $S(\tau_1;\mu,\phi;\mu',\phi')$ and $T(\tau_1;\mu,\phi;\mu',\phi')$ are,
\begin{equation}\label{eq: boundary condition1}
    I(0,-\mu,\phi) = 0\hspace{1cm} \text{and} \hspace{1cm} I(\tau_1,+\mu,\phi) = 0
\end{equation}

\cite{chandrasekhar1960radiative} used these boundary conditions in eqn.\eqref{eq: radiative transfer} and derived the four invariance principles \eqref{eq: Principle I}|\eqref{eq: Principle IV}  in terms of source functions $\xi(0,\mu,\phi)$ and $\xi(\tau_1,\mu,\phi)$ . We directly use those relations in this paper.

\section{The general integral equations for Scattering and Thermally Emitting Atmosphere}\label{sec: general integral equations}
When there is atmospheric emission as well as scattering, the source function $\xi$ can be written as \citep{sengupta2021effects}, 
\begin{equation}\label{eq: source function with emission}
\begin{split}
\xi(\tau,\mu,\phi) = \beta(\tau,\mu,\phi)+\frac{1}{4}Fe^{-\tau/\mu_0}p(\mu,\phi;-\mu_0,\phi_0) + \frac{1}{4\pi}\int_{-1}^1\int_0^{2\pi} p(\mu,\phi;\mu'';\phi'')I(\tau,\mu'',\phi'')d\phi''d\mu''
\end{split}
\end{equation}
Here $p(\mu,\phi;\mu'';\phi'')$ and $\beta(\tau,\mu,\phi)$ are the phase function and atmospheric emission respectively. This atmospheric emission $\beta$ can be expanded \citep{bellman1967chandrasekhar,sengupta2021effects} as follows,

\begin{equation}\label{eq: general emission term}
\beta(\tau;\mu;\phi) = \sum_{m=0}^{N} \beta^m (\tau,\mu)\cos m(\phi-\phi_0)
\end{equation}

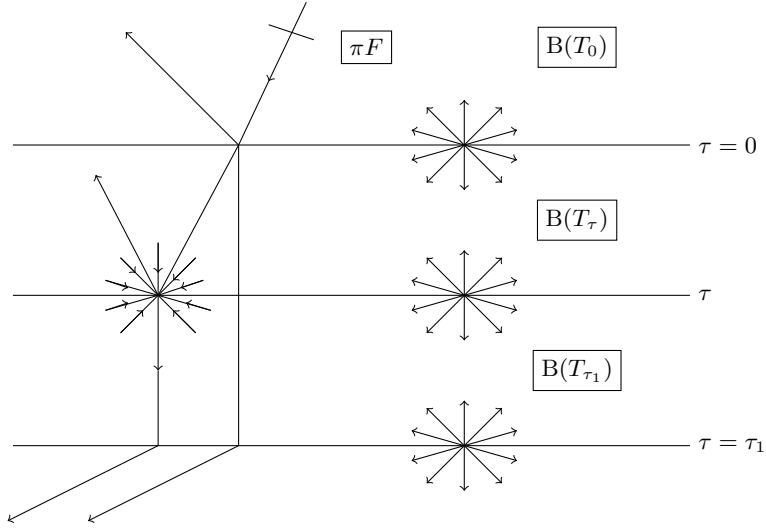
\begin{figure}\label{drawing source function}
\begin{center}
\begin{tikzpicture}
\draw (0,0)--(9,0)node[right] {$\tau=0$};
\draw (0,-2)--(9,-2)node[right]{$\tau$};
\draw (0,-4)--(9,-4)node[right]{$\tau=\tau_1$};
\draw[->] (3.9,1.9)--(3.4,0.85);
\draw (3.4,1.6)--(4.0,1.4);
\draw (3.4,0.85)--(3,0);
\draw (3,0)--(1.93,-2);
\draw[<-] (1.5,1.5)--(3,0);
\node[draw] at (4.7,1.3) {$\pi F$};
\draw (3,0) -- (3,-4);
\draw[->] (3,-4) -- (1,-5);
\draw [<-](2.23,-1.9)--(2.53,-1.8);
\draw (1.93,-2)--(2.53,-1.8);
\draw [<-](2.13,-1.8)--(2.43,-1.5);
\draw (1.93,-2)--(2.43,-1.5);
\draw [<-](1.93,-1.7)--(1.93,-1.3);
\draw (1.93,-2)--(1.93,-1.3);
\draw [<-](1.63,-1.7)--(1.43,-1.5);
\draw (1.93,-2)--(1.43,-1.5);
\draw [<-](1.53,-1.9)--(1.23,-1.8);
\draw (1.93,-2)--(1.23,-1.8);
\draw [<-](2.3,-2.1)--(2.63,-2.2);
\draw (1.93,-2)--(2.63,-2.2);
\draw [<-](2.13,-2.2)--(2.43,-2.5);
\draw (1.93,-2)--(2.43,-2.5);
\draw [<-](1.73,-2.2)--(1.43,-2.5);
\draw (1.93,-2)--(1.43,-2.5);
\draw [<-](1.53,-2.1)--(1.23,-2.2);
\draw (1.93,-2)--(1.23,-2.2);
\draw[->] (1.93,-2)--(1.1,-0.4);

\draw [->](1.93,-2)--(1.93,-3);
\draw (1.93,-3)--(1.93,-4);
\draw[->] (1.93,-4) -- (-0.07,-5);
\draw[<-] (6.7,0.2)--(6,0);
\draw[<-] (6.5,0.5)--(6,0);
\draw[<-] (6,0.6)--(6,0);
\draw[<-] (5.5,0.5)--(6,0);
\draw[<-] (5.3,0.2)--(6,0);
\draw[<-] (6.7,-0.2)--(6,0);
\draw[<-] (6.5,-0.5)--(6,0);
\draw[<-] (6,-0.6)--(6,0);
\draw[<-] (5.5,-0.5)--(6,0);
\draw[<-] (5.3,-0.2)--(6,0);
\node[draw] at (7.5,1.3) {B($T_0$)};
\draw[->] (6,-2)--(6.7,-1.8);
\draw[->] (6,-2)--(6.5,-1.5);
\draw[->] (6,-2)--(6,-1.4);
\draw[->] (6,-2)--(5.5,-1.5);
\draw[->] (6,-2)--(5.3,-1.8);
\draw[->] (6,-2)--(6.7,-2.2);
\draw[->] (6,-2)--(6.5,-2.5);
\draw[->] (6,-2)--(6,-2.6);
\draw[->] (6,-2)--(5.5,-2.5);
\draw[->] (6,-2)--(5.3,-2.2);
\node[draw] at (7.5,-1) {B($T_\tau$)};
\draw[->] (6,-4)--(6.7,-3.8);
\draw[->] (6,-4)--(6.5,-3.5);
\draw[->] (6,-4)--(6,-3.4);
\draw[->] (6,-4)--(5.5,-3.5);
\draw[->] (6,-4)--(5.3,-3.8);
\draw[->] (6,-4)--(6.7,-4.2);
\draw[->] (6,-4)--(6.5,-4.5);
\draw[->] (6,-4)--(6,-4.6);
\draw[->] (6,-4)--(5.5,-4.5);
\draw[->] (6,-4)--(5.3,-4.2);
\node[draw] at (7.5,-3) {B($T_{\tau_1}$)};
\end{tikzpicture}
\end{center}
\caption{This figure depicts the total effect of diffuse scattering and thermal emission B for \textit{finite atmosphere}. Considering an arbitrary layer with optical depth $\tau$ ($0<\tau<\tau_1$ ) we show the effect of scattering and emission, while \textit{B} is Isotropic in nature and depends only on the temperature of the emitting layer.}
\label{fig: scattering and thermal emission in finite atmosphere}
\end{figure}

For planetary atmosphere, the emission can be caused by different mechanisms (for example, see \cite{bellman1967chandrasekhar,chakrabarty2020effects,malkevich1963angular,sengupta2021effects,seager2010exoplanet}). In the current study, we consider an atmosphere, where each horizontal layer is in Local Thermodynamic Equilibrium and emits only in terms of Planck Emission \citep{seager2010exoplanet}, as shown in figure.~\ref{fig: scattering and thermal emission in finite atmosphere}. Hence, considering m=0 with no $\mu$ dependencies, eqn.\eqref{eq: general emission term} will reduce into
\begin{equation}\label{eq: thermal emission term}
\beta(\tau,\mu,\phi) \approx B(T_\tau)
\end{equation}

Here $T_\tau$ represents the absolute temperature of that particular atmospheric layer which has an optical depth $\tau$.
It is worth noting that in case of thermal emission the exact expression of $\beta$ is $\frac{\kappa}{\chi}B(T_\tau)$. But in case of low scattering limit (i.e. $\kappa >> \sigma $), the $\kappa \approx \chi$ and eqn. \eqref{eq: thermal emission term} is valid \citep{sengupta2021effects}. 

Assuming low scattering approximation, eqn.\eqref{eq: source function with emission} will become,
\begin{equation}\label{eq: source function at tau=0}
\begin{split}
\xi(0,\mu,\phi) = B(T_0)+\frac{1}{4}F[p(\mu,\phi;-\mu_0,\phi_0)
 + \frac{1}{4\pi}\int_0^1\int_0^{2\pi} p(\mu,\phi;\mu'';\phi'')S(\tau_1;\mu'',\phi'';\mu_0,\phi_0)d\phi''\frac{d\mu''}{\mu''}]
\end{split}
\end{equation}

\begin{equation}\label{eq: source function at tau=tau_1}
\begin{split}
\xi(\tau_1,\mu,\phi) = B(T_{\tau_1})+\frac{1}{4}F[e^{-\tau_1/\mu_0}p(\mu,\phi;-\mu_0,\phi_0) + \frac{1}{4\pi}\int_0^1\int_0^{2\pi} p(\mu,\phi;-\mu'';\phi'')T(\tau_1;\mu'',\phi'';\mu_0,\phi_0)d\phi''\frac{d\mu''}{\mu''}]
\end{split}
\end{equation}

Now using eqns. \eqref{eq: source function at tau=0} and \eqref{eq: source function at tau=tau_1} in the eqns. 23|26; pg:168 in \cite{chandrasekhar1960radiative} we will get (for a detailed derivation see Appendix~\ref{apndx sec: Derivation of scattering function} ) ,
\begin{equation}\label{eq: finite atmosphere scattering function1.1}
\begin{split}
&[(\frac{1}{\mu}+\frac{1}{\mu_0})S(\tau_1;\mu,\phi;\mu_0,\phi_0)+\frac{\partial S(\tau_1;\mu,\phi;\mu_0,\phi_0)}{\partial \tau_1}] =
4U(T_0)[1+\frac{1}{4\pi}\int_0^1\int_0^{2\pi} S(\tau_1;\mu,\phi;\mu',\phi')\frac{d\mu'}{\mu'}d\phi']\\
&+
p(\mu,\phi;-\mu_0,\phi_0)+ \frac{1}{4\pi}\int_0^1\int_0^{2\pi} p(\mu,\phi;\mu'';\phi'')S(\tau_1;\mu'',\phi'';\mu_0,\phi_0)d\phi''\frac{d\mu''}{\mu''}\\
&+
\frac{1}{4\pi}\int_0^1\int_0^{2\pi} S(\tau_1;\mu,\phi;\mu',\phi')[p(-\mu',\phi';-\mu_0,\phi_0)+
\frac{1}{4\pi}\int_0^1\int_0^{2\pi} p(-\mu',\phi';\mu'';\phi'')S(\tau_1;\mu'',\phi'';\mu_0,\phi_0)d\phi''\frac{d\mu''}{\mu''}]\frac{d\mu'}{\mu'}d\phi'
\end{split}
\end{equation}

\begin{equation}\label{eq: finite atmosphere scattering function2.1}
\begin{split}
&[\frac{\partial S(\tau_1;\mu,\phi;\mu_0,\phi_0)}{\partial \tau_1}] =
4U(T_{\tau_1})[e^{-\tau_1/\mu}+\frac{1}{4\pi}\int_0^1\int_0^{2\pi} T(\tau_1;\mu,\phi;\mu',\phi')\frac{d\mu'}{\mu'}d\phi']\\
&+
[exp\{-\tau_1(\frac{1}{\mu_0}+\frac{1}{\mu})\}p(\mu,\phi;-\mu_0,\phi_0) +
\frac{1}{4\pi}e^{-\tau_1/\mu}\int_0^1\int_0^{2\pi} p(\mu,\phi;-\mu'';\phi'')T(\tau_1;\mu'',\phi'';\mu_0,\phi_0)d\phi''\frac{d\mu''}{\mu''}]\\
&+
\frac{1}{4\pi}\int_0^1\int_0^{2\pi} T(\tau_1;\mu,\phi;\mu',\phi')[e^{-\tau_1/\mu_0}p(\mu',\phi';-\mu_0,\phi_0) +
\frac{1}{4\pi}\int_0^1\int_0^{2\pi} p(\mu',\phi';-\mu'';\phi'')T(\tau_1;\mu'',\phi'';\mu_0,\phi_0)d\phi''\frac{d\mu''}{\mu''}]\frac{d\mu'}{\mu'}d\phi'
\end{split}
\end{equation}

\begin{equation}\label{eq: finite atmosphere transmission function1.1}
\begin{split}
&[\frac{1}{\mu}T(\tau_1;\mu,\phi;\mu_0,\phi_0) + \frac{\partial T(\tau_1;\mu,\phi;\mu_0,\phi_0)}{\partial \tau_1}] =
4U(T_{\tau_1})[1+\frac{1}{4\pi}\int_0^1\int_0^{2\pi} S(\tau_1;\mu,\phi;\mu',\phi')\frac{d\mu'}{\mu'}d\phi']\\
&+ 
[e^{-\tau_1/\mu_0}p(-\mu,\phi;-\mu_0,\phi_0) +
\frac{1}{4\pi}\int_0^1\int_0^{2\pi} p(-\mu,\phi;-\mu'';\phi'')T(\tau_1;\mu'',\phi'';\mu_0,\phi_0)d\phi''\frac{d\mu''}{\mu''}]\\
&+
\frac{1}{4\pi}\int_0^1\int_0^{2\pi} S(\tau_1;\mu,\phi;\mu',\phi')[e^{-\tau_1/\mu_0}p(\mu',\phi';-\mu_0,\phi_0) + \frac{1}{4\pi}\int_0^1\int_0^{2\pi} p(\mu',\phi';-\mu'';\phi'')T(\tau_1;\mu'',\phi'';\mu_0,\phi_0)d\phi''\frac{d\mu''}{\mu''}] \frac{d\mu'}{\mu'}d\phi'
\end{split}
\end{equation}

\begin{equation}\label{eq: finite atmosphere transmission function2.1}
\begin{split}
&[\frac{1}{\mu_0}T(\tau_1;\mu,\phi;\mu_0,\phi_0) + \frac{\partial T(\tau_1;\mu,\phi;\mu_0,\phi_0)}{\partial \tau_1}] =
4U(T_0)[e^{-\tau_1/\mu}+\frac{1}{4\pi}\int_0^1\int_0^{2\pi} T(\tau_1;\mu,\phi;\mu',\phi')\frac{d\mu'}{\mu'}d\phi']\\
&+
[p(-\mu,\phi;-\mu_0,\phi_0) + \frac{1}{4\pi}\int_0^1\int_0^{2\pi} p(-\mu,\phi;\mu'';\phi'')S(\tau_1;\mu'',\phi'';\mu_0,\phi_0)d\phi''\frac{d\mu''}{\mu''}]e^{-\tau_1/\mu}\\
&+
\frac{1}{4\pi}\int_0^1\int_0^{2\pi} T(\tau_1;\mu,\phi;\mu',\phi')[p(-\mu',\phi';-\mu_0,\phi_0) + \frac{1}{4\pi}\int_0^1\int_0^{2\pi} p(-\mu',\phi';\mu'',\phi'')S(\tau_1;\mu'',\phi'';\mu_0,\phi_0)d\phi''\frac{d\mu''}{\mu''}]\frac{d\mu'}{\mu'}d\phi'
\end{split}
\end{equation}
where, $U(T_0)=\frac{B(T_0)}{F}$, $U(T_{\tau_1})=\frac{B(T_{\tau_1})}{F}$.

Equations. \eqref{eq: finite atmosphere scattering function1.1} | \eqref{eq: finite atmosphere transmission function2.1} represent the integral equations governing the problem of diffuse reflection and trasmission in presence of atmospheric thermal emission of a plane-parallel atmosphere with finite optical depth.
\section{The integral equations in isotropic scattering}\label{sec: specific form of integral equations}
It is evident that these four integral equations have an explicit dependency on the phase function $p(\mu,\phi;\mu',\phi')$. The different types of phase functions are discussed in \cite{chandrasekhar1960radiative,sengupta2021effects}. Here, we specifically study the effect of thermal emission in isotropic scattering case only. It can be treated in terms of single scattering albedo $\tilde{\omega}_0$ \citep{sengupta2021effects} as,
$$p(\mu,\phi;\mu_0,\phi_0)=\tilde{\omega_0}$$

This axial symmetry in phase function is also a property of scattering and transmission functions and they can be expressed in axisymmetric terms as, $S(\tau_1;\mu;\mu')$ and $T(\tau_1;\mu;\mu')$

Then, eqn.\eqref{eq: finite atmosphere scattering function2.1} will become,
\begin{equation*}
\begin{split}
&[\frac{\partial S(\tau_1;\mu,\mu_0)}{\partial \tau_1}]=
4U(T_{\tau_1})[e^{-\tau_1/\mu}+\frac{1}{2}\int_0^1 T(\tau_1;\mu;\mu')\frac{d\mu'}{\mu'}]\\
&+
[exp\{-\tau_1(\frac{1}{\mu_0}+\frac{1}{\mu})\}\tilde{\omega_0}+
\frac{1}{2}e^{-\tau_1/\mu}\int_0^1 \tilde{\omega_0}T(\tau_1;\mu'',\mu_0)\frac{d\mu''}{\mu''}] +
\frac{1}{2}\int_0^1 T(\tau_1;\mu,\mu')[e^{-\tau_1/\mu_0}\tilde{\omega_0}+\frac{1}{2}\int_0^1 \tilde{\omega_0}T(\tau_1;\mu'',\mu_0)\frac{d\mu''}{\mu''}]\frac{d\mu'}{\mu'}
\end{split}
\end{equation*}
\begin{equation}\label{eq: finite atmosphere isotropic scattering function1}
\boxed{
[\frac{\partial S(\tau_1;\mu,\mu_0)}{\partial \tau_1}]=4U(T_{\tau_1})W(\mu)+\tilde{\omega_0}W(\mu_0)W(\mu)
}
\end{equation}
Here we define two new functions as,
\begin{equation}\label{eq: V-function 1}
\begin{split}
V(\mu)&=
1+\frac{1}{2}\int_0^1 S(\tau_1;\mu,\mu')\frac{d\mu'}{\mu'}
\end{split}
\end{equation}
and
\begin{equation}\label{eq: W-function 1}
\begin{split}
W(\mu)&=
e^{-\tau_1/\mu}+\frac{1}{2}\int_0^1 T(\tau_1;\mu,\mu')\frac{d\mu'}{\mu'}
\end{split}
\end{equation}

Similarly, eqns.\eqref{eq: finite atmosphere scattering function1.1}, \eqref{eq: finite atmosphere transmission function1.1}, \eqref{eq: finite atmosphere transmission function2.1} can be expressed in terms of V and W functions as follows,

\begin{equation}\label{eq: final scattering function}
\boxed{
(\frac{1}{\mu}+\frac{1}{\mu_0})S(\tau_1;\mu,\mu_0)=
4[U(T_0)V(\mu)-U(T_{\tau_1})W(\mu)]+\tilde{\omega_0[}V(\mu)V(\mu_0)-W(\mu)W(\mu_0)]
}
\end{equation}

\begin{equation}\label{eq: finite atmosphere isotropic transmission function1}
\begin{split}
[\frac{1}{\mu_0}T(\tau_1;\mu,\mu_0) + \frac{\partial T(\tau_1;\mu,\mu_0)}{\partial \tau_1}] 
=
4U(T_0)W(\mu)+\tilde{\omega_0}V(\mu_0)W(\mu)
\end{split}
\end{equation}

\begin{equation}\label{eq: finite atmosphere isotropic transmission function2}
\begin{split}
[\frac{1}{\mu}T(\tau_1;\mu,\mu_0) + \frac{\partial T(\tau_1;\mu,\mu_0)}{\partial \tau_1}] =
4U(T_{\tau_1})V(\mu)+\tilde{\omega_0}W(\mu_0)V(\mu)
\end{split}
\end{equation}

Subtracting eqn.\eqref{eq: finite atmosphere isotropic transmission function2} from eqn.\eqref{eq: finite atmosphere isotropic transmission function1} gives,
\begin{equation}\label{eq: finite atmosphere isotropic transmission function3}
\boxed{
(\frac{1}{\mu_0}-\frac{1}{\mu})T(\tau_1;\mu,\mu_0)= 4[U(T_0)W(\mu)-U(T_{\tau_1})V(\mu)]+\tilde{\omega_0}[V(\mu_0)W(\mu)-W(\mu_0)V(\mu)]
}
\end{equation}

Subtracting eqn.\eqref{eq: finite atmosphere isotropic transmission function1}*$\frac{1}{\mu}$ from eqn.\eqref{eq: finite atmosphere isotropic transmission function2}*$\frac{1}{\mu_0}$ gives,
\begin{equation}\label{eq: finite atmosphere isotropic transmission function4}
\boxed{
(\frac{1}{\mu_0}-\frac{1}{\mu})\frac{\partial T(\tau_1;\mu,\mu_0)}{\partial \tau_1}=4[\frac{1}{\mu_0}U(T_{\tau_1})V(\mu)-\frac{1}{\mu}U(T_0)W(\mu)]+\tilde{\omega_0}[\frac{1}{\mu_0}W(\mu_0)V(\mu)-
\frac{1}{\mu}V(\mu_0)W(\mu)]
}
\end{equation}

Thus, the functional form of V and W functions (eqns.\eqref{eq: V-function 1} and \eqref{eq: W-function 1}) can be modified as,
\begin{equation}\label{eq: V-function 2}
\begin{split}
V(\mu)=
1+\frac{1}{2}\mu\int_0^1 \{4[U(T_0)V(\mu)-U(T_{\tau_1})W(\mu)]+\tilde{\omega_0[}V(\mu)V(\mu_0)-W(\mu)W(\mu_0)]\}\frac{d\mu'}{\mu'+\mu}
\end{split}
\end{equation}
and
\begin{equation}\label{eq: W-function 2}
\begin{split}
W(\mu)=e^{-\tau_1/\mu}+\frac{1}{2}\mu\int_0^1 4[U(T_0)W(\mu)-4U(T_{\tau_1})V(\mu)]+\tilde{\omega_0}[V(\mu_0)W(\mu)-W(\mu_0)V(\mu)]\frac{d\mu'}{\mu-\mu'}
\end{split}
\end{equation}

The final emitted radiation from $\tau=0$ and $\tau=\tau_1$ can be expressed from eqn. \eqref{eq: diffuse reflection1} as,
\begin{equation}\label{eq: final radiation from tau=0}
\begin{split}
I(0,\mu) =
\frac{\mu_0}{\mu+\mu_0}
[B(T_0)V(\mu)-B(T_{\tau_1})W(\mu)]+
\frac{F}{4}\frac{\mu_0}{\mu+\mu_0}\tilde{\omega_0}[V(\mu)V(\mu_0)-W(\mu)W(\mu_0)]
\end{split}
\end{equation}
and,
\begin{equation}\label{eq: final radiation from tau=tau1}
\begin{split}
I(\tau_1,-\mu)=\frac{\mu_0}{\mu-\mu_0}[B(T_0)W(\mu)-B(T_{\tau_1})V(\mu)]+\frac{F}{4}\frac{\mu_0}{\mu-\mu_0}\tilde{\omega_0}[V(\mu_0)W(\mu)-W(\mu_0)V(\mu)]
\end{split}
\end{equation}

Eqns.\eqref{eq: final radiation from tau=0} and \eqref{eq: final radiation from tau=tau1} can be expressed as,
\begin{equation}\label{eq: Invariance principle matrix form1}
\begin{split}
\begin{bmatrix}
(\mu+\mu_0)I(0,+\mu)\\
(\mu-\mu_0)I(\tau_1,-\mu)
\end{bmatrix}
&=
\begin{bmatrix}
V(\mu) & W(\mu)\\
W(\mu) & V(\mu)
\end{bmatrix}
\begin{bmatrix}
\frac{F}{4}\tilde{\omega_0}\mu_0V(\mu_0)+\mu_0B(T_0)\\
-\frac{F}{4}\tilde{\omega_0}\mu_0W(\mu_0)-\mu_0B(T_{\tau_1})
\end{bmatrix}
\end{split}
\end{equation}

\section{A simple transformation rule}\label{sec: transformation rule}

\cite{chandrasekhar1960radiative} introduced two crucial functions $S(\tau_1;\mu,\phi;\mu_0,\phi_0)$ and $T(\tau_1;\mu,\phi;\mu_0,\phi_0)$ while considering diffuse scattering in finite atmosphere. The transformation rule between these two functions was established by \cite{coakley1973simple} as,
\begin{equation}\label{eq: S-T Transformation rule}
\begin{split}
    &S(\tau_1,\mu,\phi;-\mu_0,\phi_0)e^{-\tau_1/\mu_0} = T(\tau_1,\mu,\phi;\mu_0,\phi_0)\\
    &T(\tau_1,\mu,\phi;-\mu_0,\phi_0)e^{-\tau_1/\mu_0} = S(\tau_1,\mu,\phi;\mu_0,\phi_0)
    \end{split}
\end{equation}

Here we will show that these rules are indeed true while the thermal emission is included in this problem under some circumstances. We replace $\mu_0$ by $-\mu_0$ in eqn.\eqref{eq: finite atmosphere scattering function1.1} and multiplying both sides by $e^{-\tau_1/\mu_0}$ and get,
\begin{equation}\label{eq: finite atmosphere scattering function1.1.1}
\begin{split}
\therefore
&[(\frac{1}{\mu}-\frac{1}{\mu_0})S(\tau_1;\mu,\phi;-\mu_0,\phi_0)e^{-\tau_1/\mu_0} + \frac{\partial S(\tau_1;\mu,\phi;-\mu_0,\phi_0)}{\partial \tau_1} e^{-\tau_1/\mu_0}]\\
&=
4U(T_0)[1+\frac{1}{4\pi}\int_0^1\int_0^{2\pi} S(\tau_1;\mu,\phi;\mu',\phi')\frac{d\mu'}{\mu'}d\phi']e^{-\tau_1/\mu_0}\\
&+
p(\mu,\phi;\mu_0,\phi_0)e^{-\tau_1/\mu_0} + \frac{1}{4\pi}\int_0^1\int_0^{2\pi} p(\mu,\phi;\mu'';\phi'')S(\tau_1;\mu'',\phi'';-\mu_0,\phi_0) e^{-\tau_1/\mu_0} d\phi''\frac{d\mu''}{\mu''}\\
&+
\frac{1}{4\pi}\int_0^1\int_0^{2\pi} S(\tau_1;\mu,\phi;\mu',\phi')[p(-\mu',\phi';\mu_0,\phi_0)e^{-\tau_1/\mu_0}\\
&+
\frac{1}{4\pi}\int_0^1\int_0^{2\pi} p(-\mu',\phi';\mu'';\phi'')S(\tau_1;\mu'',\phi'';-\mu_0,\phi_0)e^{-\tau_1/\mu_0}d\phi''\frac{d\mu''}{\mu''}]\frac{d\mu'}{\mu'}d\phi'
\end{split}
\end{equation}

Now if we make use of eqn.\eqref{eq: S-T Transformation rule} and the symmetric properties of the phase function  $p(\mu,\phi;-\mu_0,\phi_0) = p(-\mu,\phi;\mu_0,\phi_0)$ and $p(-\mu,\phi;-\mu_0,\phi_0) = p(\mu,\phi;\mu_0,\phi_0)$ then equation \eqref{eq: finite atmosphere scattering function1.1.1} will be,
\begin{equation}\label{eq: finite atmosphere scattering function1.1.2}
\begin{split}
\therefore
&\frac{1}{\mu}T(\tau_1;\mu,\phi;\mu_0,\phi_0) + \frac{\partial T(\tau_1;\mu,\phi;\mu_0,\phi_0)}{\partial \tau_1}=
4U(T_0)e^{-\tau_1/\mu_0}[1+\frac{1}{4\pi}\int_0^1\int_0^{2\pi} S(\tau_1;\mu,\phi;\mu',\phi')\frac{d\mu'}{\mu'}d\phi']\\
&+
p(-\mu,\phi;-\mu_0,\phi_0)e^{-\tau_1/\mu_0} + \frac{1}{4\pi}\int_0^1\int_0^{2\pi} p(\mu,\phi;\mu'';\phi'')T(\tau_1;\mu'',\phi'';\mu_0,\phi_0) d\phi''\frac{d\mu''}{\mu''}\\
&+
\frac{1}{4\pi}\int_0^1\int_0^{2\pi} S(\tau_1;\mu,\phi;\mu',\phi')[p(\mu',\phi';-\mu_0,\phi_0)e^{-\tau_1/\mu_0}\\
&+
\frac{1}{4\pi}\int_0^1\int_0^{2\pi} p(\mu',\phi';-\mu'';\phi'')T(\tau_1;\mu'',\phi'';\mu_0,\phi_0)d\phi''\frac{d\mu''}{\mu''}]\frac{d\mu'}{\mu'}d\phi'
\end{split}
\end{equation}

Observing the similarity of this equation with eqn.\eqref{eq: finite atmosphere transmission function1.1}, we can write condition on thermal emission,
\begin{equation}\label{eq: reduced thermal emission}
    U(T_{\tau_1}) = U(T_0) e^{-\tau_1/\mu_0} \hspace{1cm}  \hspace{1cm}B(T_{\tau_1}) = B(T_0) e^{-\tau_1/\mu_0}
\end{equation}

So the transformation rule between S- and T-function eqn.\eqref{eq: S-T Transformation rule} will be valid when the thermal emission from different atmospheric layers are connected by eqn.\eqref{eq: reduced thermal emission}. This blackbody emission can be named as the \textit{reduced thermal emission}.

For isotropic scattering case the transformation rules of V and W can be obtained using eqns.\eqref{eq: S-T Transformation rule}, \eqref{eq: V-function 1}, \eqref{eq: W-function 1} as,
\begin{equation}\label{eq: V-W transformation rule}
    V(-\mu)e^{-\tau_1/\mu} = W(\mu) \hspace{1cm}  \hspace{1cm} W(-\mu)e^{-\tau_1/\mu} = V(\mu)
\end{equation}

\section{The Physical meaning of V and W functions:}\label{sec: physical meaning}

We introduced two new functions $V(\mu)$ and $W(\mu)$ which resembles to Chandrasekhar's $X(\mu)$- and $Y(\mu)$- functions respectively in the presence of thermal emission with diffusely reflecting finite atmosphere problem . The physical meaning of X- and Y- functions has been discussed in \cite{chandrasekhar1960radiative,van1948scattering,peraiah2002an}. Here we will discuss the additional effects introduced in V and W functions.

\begin{figure}
\begin{center}
\begin{tikzpicture}
\filldraw[color=black!100, fill=red!0, very thick](6,-1) circle (4);
\draw[->] (8,3)--(9,4)node[right] {V($\mu$)};
\draw[->] (9,-4)--(10,-4.5)node[right] {W($\mu$)};
\filldraw[color=black!100, fill=black!0, very thick](6,1.5) circle (0.1);
\draw[<-] (6,2.7)--(6,1.6);
\draw[<-] (3,0)--(5.9,1.4);
\draw[<-] (9,0)--(6.1,1.4);
\draw[<-] (6.7,2.7)--(6.1,1.6);
\draw[<-] (7.1,2.2)--(6.1,1.5);
\draw[<-] (5.3,2.7)--(5.9,1.6);
\draw[<-] (5,2.1)--(5.9,1.5);
\draw[<-] (7.5,1.5)--(6.1,1.45);
\draw[<-] (4.5,1.45)--(5.9,1.45);
\draw[<-] (6,1)--(6,1.4);
\draw[<-] (6.3,1)--(6,1.4);
\draw[<-] (5.7,1)--(6,1.4);
\draw (0,0)--(12,0)node[right] {$\tau=0$};
\draw (0,-2)--(12,-2)node[right]{$\tau=\tau_1$};
\draw[<-] (3.7,0.2)--(3,0);
\draw[<-] (3.5,0.5)--(3,0);
\draw[<-] (3,0.6)--(3,0);
\draw[<-] (2.5,0.5)--(3,0);
\draw[<-] (2.3,0.2)--(3,0);
\draw[<-] (6.7,0.2)--(6,0);
\draw[<-] (6.5,0.5)--(6,0);
\draw[<-] (6,0.6)--(6,0);
\draw[<-] (5.5,0.5)--(6,0);
\draw[<-] (5.3,0.2)--(6,0);
\draw[<-] (6.7,-0.2)--(6,0);
\draw[<-] (6.5,-0.5)--(6,0);
\draw[<-] (6,-0.6)--(6,0);
\draw[<-] (5.5,-0.5)--(6,0);
\draw[<-] (5.3,-0.2)--(6,0);
\node[draw] at (7.6,-0.8) {B($T_0$)};
\draw[<-] (9.7,0.2)--(9,0);
\draw[<-] (9.5,0.5)--(9,0);
\draw[<-] (9,0.6)--(9,0);
\draw[<-] (8.5,0.5)--(9,0);
\draw[<-] (8.3,0.2)--(9,0);

\draw[->] (3,-2)--(3.7,-2.2);
\draw[->] (3,-2)--(3.5,-2.5);
\draw[->] (3,-2)--(3,-2.6);
\draw[->] (3,-2)--(2.5,-2.5);
\draw[->] (3,-2)--(2.3,-2.2);
\draw[->] (6,-2)--(6.7,-1.8);
\draw[->] (6,-2)--(6.5,-1.5);
\draw[->] (6,-2)--(6,-1.4);
\draw[->] (6,-2)--(5.5,-1.5);
\draw[->] (6,-2)--(5.3,-1.8);
\draw[->] (6,-2)--(6.7,-2.2);
\draw[->] (6,-2)--(6.5,-2.5);
\draw[->] (6,-2)--(6,-2.6);
\draw[->] (6,-2)--(5.5,-2.5);
\draw[->] (6,-2)--(5.3,-2.2);
\node[draw] at (6.1,-3) {B$(T_{\tau_1})$};
\draw[->] (9,-2)--(9.7,-2.2);
\draw[->] (9,-2)--(9.5,-2.5);
\draw[->] (9,-2)--(9,-2.6);
\draw[->] (9,-2)--(8.5,-2.5);
\draw[->] (9,-2)--(8.3,-2.2);
\end{tikzpicture}
\end{center}
\caption{This figure explains the physical meaning of $V(\mu)$ and $W(\mu)$ functions. The total brightness observed by a distant observer will be $V(\mu)$ (for the observer above $\tau=0$) and $W(\mu)$ (for the observer below $\tau=\tau_1$) times the brightness of the point source alone. Hence all the scattering, transmission and thermal emission from the atmosphere contributes in terms of V and W functions.}
\label{fig: V_mu and W_mu explanation}
\end{figure}
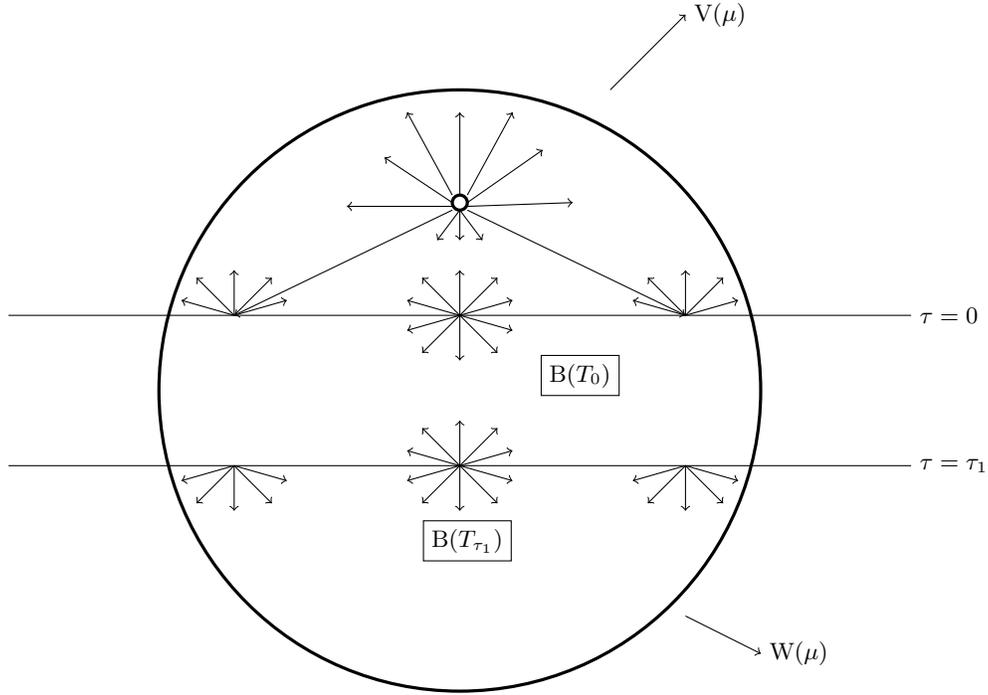

Let there be a point source above the layer $\tau=0$ which has unit brightness (with the total emitting flux $4\pi$ from the point source). Now the flux will be scattered and transmitted multiple times by both the atmospheric layers at $\tau=0$ and $\tau =\tau_1$ (see fig~\ref{fig: V_mu and W_mu explanation}). In addition to it there is a contribution of thermal emission $B(T_0)$ and $B(T_{\tau_1})$ respectively from those layers. Now for an observer at large distance, the combination of the same point source and the illuminated atmosphere will again appear as a point source and only the combined effect can be observed. If that distant observer is in $(+\mu,\phi)$ direction from the atmosphere (i.e. above the atmospheric layer $\tau =0$ in fig.~\ref{fig: V_mu and W_mu explanation}), then $V(\mu)$ will be the total observed brightness. In the same way, if the observer is in $(-\mu,\phi)$ direction from the atmosphere (i.e. below the atmospheric layer $\tau =\tau_1$ in fig.~\ref{fig: V_mu and W_mu explanation}), then $W(\mu)$ will be the total observed brightness. In both the cases, the factor $\frac{1}{\mu}$ is positive.

In other words,  $V(\mu)$ and $W(\mu)$ represents the relative change of the incident and transmitted flux along $(+\mu)$ and $(-\mu)$ direction respectively due to the presence of the atmosphere. This relative change shows the combined effect of scattering, transmission and thermal emission by the atmospheric layers.

Clearly in the absence of thermal emission, the contribution of thermal emission will be removed and the observed brightness will be a combination of atmospheric scattering and transmission of the point source flux only. In such circumstances, $V(\mu)$ and $W(\mu)$ functions will reduce into the Chandrasekhar's $X(\mu)$ and $Y(\mu)$ functions only (see section~\ref{sec: consistency} for more discussion). Also figure~\ref{fig: V_mu and W_mu explanation} will reduce into the figure given in \cite{van1948scattering}.

\begin{figure}
    \begin{center}
        \begin{tikzpicture}
        \draw (10,0) arc(10:170:4cm);
        \draw[->] (8,3)--(9,4) node[right]{$M(\mu)$};
        \filldraw[color=black!100, fill=black!0, very thick](6,1.5) circle (0.1);
        \draw[<-] (6,2.7)--(6,1.6);
\draw[<-] (3,0)--(5.9,1.4);
\draw[<-] (9,0)--(6.1,1.4);
\draw[<-] (6.7,2.7)--(6.1,1.6);
\draw[<-] (7.1,2.2)--(6.1,1.5);
\draw[<-] (5.3,2.7)--(5.9,1.6);
\draw[<-] (5,2.1)--(5.9,1.5);
\draw[<-] (7.5,1.5)--(6.1,1.45);
\draw[<-] (4.5,1.45)--(5.9,1.45);
\draw[<-] (6,1)--(6,1.4);
\draw[<-] (6.3,1)--(6,1.4);
\draw[<-] (5.7,1)--(6,1.4);
\draw (0,0)--(12,0)node[right] {$\tau=0$};
\draw (0,-2)--(12,-2)node[right]{$\tau=\tau_1=\infty$};
\draw[<-] (3.7,0.2)--(3,0);
\draw[<-] (3.5,0.5)--(3,0);
\draw[<-] (3,0.6)--(3,0);
\draw[<-] (2.5,0.5)--(3,0);
\draw[<-] (2.3,0.2)--(3,0);
\draw[<-] (6.7,0.2)--(6,0);
\draw[<-] (6.5,0.5)--(6,0);
\draw[<-] (6,0.6)--(6,0);
\draw[<-] (5.5,0.5)--(6,0);
\draw[<-] (5.3,0.2)--(6,0);
\draw[<-] (6.7,-0.2)--(6,0);
\draw[<-] (6.5,-0.5)--(6,0);
\draw[<-] (6,-0.6)--(6,0);
\draw[<-] (5.5,-0.5)--(6,0);
\draw[<-] (5.3,-0.2)--(6,0);
\node[draw] at (7.6,-0.8) {B($T$)};
\draw[<-] (9.7,0.2)--(9,0);
\draw[<-] (9.5,0.5)--(9,0);
\draw[<-] (9,0.6)--(9,0);
\draw[<-] (8.5,0.5)--(9,0);
\draw[<-] (8.3,0.2)--(9,0);
        \end{tikzpicture}
    \end{center}
    \caption{A pictorial view of the M($\mu$)-function derived in \cite{sengupta2021effects} for semi-infinite atmosphere. The total radiation observed from $(+\mu,\phi)$ direction will be the addition of the light from the source, scattered by the atmosphere and the thermally emitted radiation from the atmospheric layer. Here all the layers are at same temperature T.}
    \label{fig: M_mu explanation}
\end{figure}
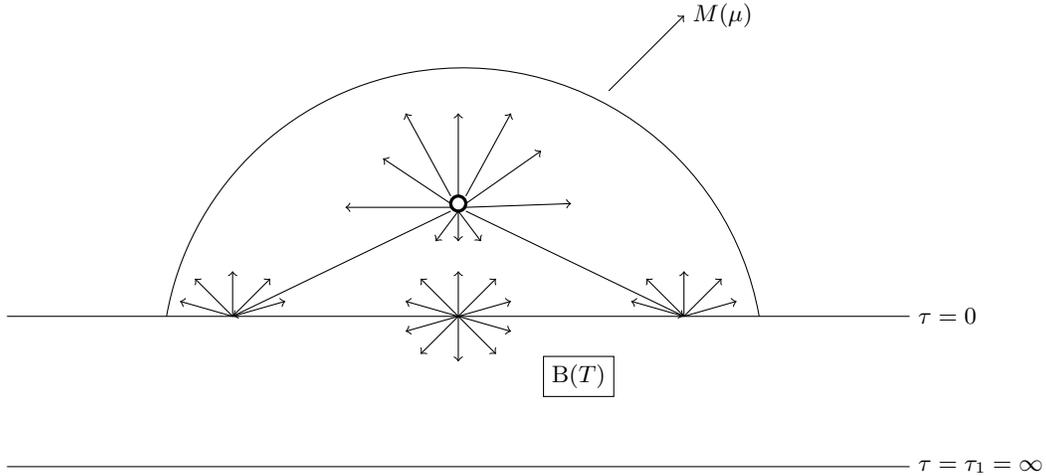

In case of semi-infinite atmosphere, the bottom layer will be extended at $\tau_1\to \infty$ as shown in figure~\ref{fig: M_mu explanation}. In such circumstances, the distant observer can observe the combined effect from $(+\mu,\phi)$ direction only. Hence $W(\mu)$ function will vanish and $V(\mu)$ function will give the combined effect of scattering and thermal emission. In such case $V(\mu)$-function will reduce into the well known $M(\mu)$-function as introduced by \cite{sengupta2021effects} for semi-infinite atmosphere case.

Hence the V and W functions represents the relative change of the flux from point source due to the presence of atmospheric scattering, transmission and thermal emission.

\section{Consistency Check}\label{sec: consistency}
In this section we will show that how our results reduce into previous literature results at specific boundary conditions. It is expected that, when the atmospheric thermal emission is very less than the incident flux (i.e. $B(T_0) , B(T_{\tau_1}) << F$) then our solutions should match with the results of only scattering case as derived by \cite{chandrasekhar1960radiative}.

In case of no thermal emission limit, $U(T_0) , U(T_{\tau_1})\to 0$ and thus equations \eqref{eq: V-function 2} and \eqref{eq: W-function 2} will reduce into that of Chandrasekhar's X and Y functions as shown in \cite{chandrasekhar1960radiative} (pg.181; eqns.84-85) 
$$\lim_{U\to 0} V(\mu)\to X(\mu)$$ and
$$\lim_{U\to 0}W(\mu)\to Y(\mu)$$.

Hence in the limit of $U\to 0$, eqn.\eqref{eq: final scattering function} and \eqref{eq: finite atmosphere isotropic transmission function3} will become,
\begin{equation}
\begin{split}
&(\frac{1}{\mu}+\frac{1}{\mu_0})S(\tau_1;\mu,\mu_0)=
\tilde{\omega_0}[X(\mu)X(\mu_0)-Y(\mu)Y(\mu_0)]\\
&\text{and}\\
& (\frac{1}{\mu_0}+\frac{1}{\mu})T(\tau_1;\mu,\mu_0)=
\tilde{\omega_0}[X(\mu_0)Y(\mu)-Y(\mu_0)X(\mu)]
\end{split}
\end{equation}

These equations are the same as given in \cite{chandrasekhar1960radiative} (pg.181; eqns. 80-81)
The no thermal emission limit will also affect the final radiation coming out from both the boundaries at $\tau=0$ and $\tau=\tau_1$. Hence equations \eqref{eq: final radiation from tau=0} and \eqref{eq: final radiation from tau=tau1} will reduce into the scattering only case. At no thermal emission case we can write the matrix equation \eqref{eq: Invariance principle matrix form1} as follows,

\begin{equation}\label{Invariance principle matrix form without emission}
\begin{split}
\begin{bmatrix}
(\mu+\mu_0)I(0,+\mu)\\
(\mu-\mu_0)I(\tau_1,-\mu)
\end{bmatrix}
&=
\begin{bmatrix}
X(\mu) & Y(\mu)\\
Y(\mu) & X(\mu)
\end{bmatrix}
\begin{bmatrix}
\frac{F}{4}\tilde{\omega_0}\mu_0X(\mu_0)\\
-\frac{F}{4}\tilde{\omega_0}\mu_0Y(\mu_0)
\end{bmatrix}
\end{split}
\end{equation}

This is the same form as derived by \citep{chandrasekhar1960radiative} (pg. 201; eqns 108 and 109) in case of diffuse scattering. 

Now we will show the two limiting cases of optical depth.
\begin{enumerate}
    \item Semi-Infinite optical depth $(\tau_1\to \infty)$:  In this condition the expression of the function $V(\mu)$ will reduced into ,
    \begin{equation}\label{eq: V at tau_1 infinity limit}
        V(\mu) = 1+2U(T)M(\mu)\mu \log(1+\frac{1}{\mu})+ \frac{\tilde{\omega_0}}{2}\mu M(\mu)\int_0^1 \frac{M(\mu')}{\mu+\mu'}d\mu'=M(\mu)
    \end{equation}
    
    This expression is same as of the M-function derived in \citep{sengupta2021effects} in the context of semi-infinite atmosphere. Hence we can say that the finite atmosphere problem boils down to semi-infinite atmosphere problem at this limiting value.
    
    Now using the transformation rule \eqref{eq: V-W transformation rule}, the W-function can be represented as, 
\begin{equation}\label{eq: W at tau_1 infinity limit}
    W(\mu) = \lim_{\tau_1\to\infty} V(-\mu)e^{-\tau_1/\mu}\to 0
\end{equation}
    \item Small optical depth $(\tau_1\to 0)$:
In such case the W function will be,
\begin{equation}\label{eq: W at tau_1 zero limit}
    W(\mu)\to e^{-\tau_1/\mu}
\end{equation}
and using the transformation rule we get,
\begin{equation}\label{eq: V at tau_1 zero limit}
    V(\mu) = W(-\mu)e^{-\tau_1/\mu} \to 1
\end{equation}

These values are same as shown in \citep{peraiah2002an} in the case of Y and X functions respectively.
\end{enumerate}


\section{Discussion}\label{sec: Discussion}

    The finite atmosphere diffuse reflection problem  was first introduced by \cite{chandrasekhar1960radiative} for scattering only atmosphere where no atmospheric emission was considered. Here for the first time we include the thermal emission effect simultaneously with the isotropic scattering from each atmospheric layers in finite atmosphere diffuse reflection problem. The thermal emission modifies Chandrasekhar's results in terms of the factor $U(T)$, where U is the ratio of blackbody emission (B) and irradiation flux (F) (see section~\ref{sec: general integral equations} and \ref{sec: specific form of integral equations}). Moreover, the modified scattering and transmission functions obey the same transformation rules as established by \cite{coakley1973simple} (as shown in sec. \ref{sec: transformation rule}). Then, we show that our results are consistent with that of the \cite{chandrasekhar1960radiative} results in the limit of low atmospheric thermal emission (i.e. $B<<F$). Hence, it can be said that our treatment of thermal emission and scattering for finite atmosphere problem is more general than Chansdrasekhar's one. In exoplanetary context Chandrasekhar's results are used to model the reflection, transmission and emission spectra for highly irradiated low emitting planets \cite{madhusudhan2012analytic}. But as it is evident that when the thermal emission and scattering occurs comparably in a planetary atmosphere (e.g. low irradiating ultra-hot jupiters), then our results will provide more accurate modeling than Chandrasekhar's results.

    The thermal emission from each atmospheric layer will travel through other layers as well and will face the scattering and transmission. For example, the emission from the layer at $\tau=0$ is $B(T_0)$ which scattered along the direction $(\mu,\phi)$ and contribute to the final radiation $I(0,\mu)$ in terms of $B(T_0)V(\mu)$ (see eqn.\eqref{eq: final radiation from tau=0}). In the same way it contributes along the direction $(-\mu,\phi)$ in the radiation in terms of $B(T_0)W(\mu)$. Thus, the flux F irradiated on the atmosphere and the atmospheric thermal emission will follow the same rule of scattering and transmission. So this theory is applicable to those planetary atmosphere where, (1) the atmospheric thermal emission is comparable to the irradiated stellar flux and (2) the atmosphere gives infrared scattering effects.

    Here we have revisited the connection relation between scattering and transmission functions (see \eqref{eq: S-T Transformation rule}) as established by \cite{coakley1973simple}. This transformation rule describes the interchange between S- and T- functions depending on the orientation of the incident beam in case of only diffuse scattering in finite atmosphere \citep{coakley1973simple}. In this work, we first show that this relation is indeed true in case of thermally emitting atmosphere. Secondly, a transformation rule for the V and W functions (see eqn. \eqref{eq: V-W transformation rule}) as well as a connection relation between the thermal emission flux at different atmospheric layers, which can be named as \textit{reduced thermal emission}, is established. It ensures that if a beam incident from the upper side of the layer $\tau=0$, then V and W functions can be represented as shown in the figure \ref{fig: V_mu and W_mu explanation}. But while the light beam incident from the lower side of $\tau = \tau_1$ layer then the position of V and W functions in figure~\ref{fig: V_mu and W_mu explanation} will interchange. It shows the symmetry of the solutions provided here. Although the transformation rules are applicable only if the atmospheric emission at different optical depths are connected by the relation given in eqn. \eqref{eq: reduced thermal emission}.
    
    The applicability of the transformation rule in case of thermal emission, also ensures the fact that in case of comparable emission and scattering from an atmosphere, these both effects are in equal footing. Hence, both effects should be considered for a full-proof modeling.

    The $V(\mu)$ and $W(\mu)$ functions are analogous to Chandrasekhar's X and Y functions mentioned in \cite{chandrasekhar1960radiative}. They represent the relative changes of the radiation from the layer $\tau = 0$ along the direction $(\mu,\phi)$ and from $\tau=\tau_1$ along $(-\mu,\phi)$ respectively due to the presence of the atmosphere. Hence, the atmospheric presence can be realized in terms of diffuse reflection and atmospheric thermal emission from the corresponding layer. In other words it can be said that they act as the source function for the direction $(\mu,\phi)$ at $\tau=0$ and the direction $(-\mu,\phi)$ at $\tau=\tau_1$ respectively.

         We showed (in section~\ref{sec: consistency}) that at the semi-infinite limit (i.e.$\tau_1\to \infty$), our finite atmosphere results will reduce into that of semi-infinite results obtained in \cite{sengupta2021effects}. Hence we can say that the $M(\mu)$-function (see \cite{sengupta2021effects} for detail) is semi-infinite counterpart of the more general $V(\mu)$-function as shown in figure~\ref{fig: M_mu explanation}. It reveals the semi-infinite limiting case of $V(\mu)$-function. 
        
         The work presented by \citep{sengupta2021effects} considered atmospheric thermal emission in semi-infinite atmosphere case and thus limited by the condition of translational invariant thermal emission in the atmosphere. For Planetary atmosphere it means that such theory is applicable only for those planets which has an isothermal atmosphere. In this work we removed that limitation by considering the finite atmosphere problem which does not need a translational invariant thermal emission and in such case the scattering function $S(\tau,\mu,\phi;\mu_0,\phi_0)$ varies with the optical depth of the atmosphere. It will provide the opportunity to model the atmospheric spectra for any type of atmospheric temperature structure with simultaneous emission and scattering.

     In this work we considered only the isotropic scattering case which is indeed the first step to modify the finite atmosphere scattering problem in the presence of thermal emission. This work can be expanded for the general cases of scattering with the same recipe and the modifications will follow accordingly. To include the numerical approach, one should use the Heyney-Greenstein phase function \cite{henyey1941diffuse},
    \begin{equation*}
p(\cos\Theta) = \frac{1-g^2}{(1+g^2-2g\cos\Theta)^\frac{2}{3}}
\end{equation*}
where, $g\in[-1,1]$ is the asymmetry parameter, as shown in \cite{bellman1967chandrasekhar,batalha2019exoplanet}. 
       
        The atmospheric thermal emission is the simplest possible emission process considered in our work. However, we assumed low scattering limit $\sigma<<\kappa$ to use the simple planck function as the atmospheric emission term (see eqn. \eqref{eq: thermal emission term}). It simplifies the mathematical derivations significantly. However, this restriction can be removed by replacing $B(T_\tau)$ with $\frac{\kappa}{\chi}B(T_\tau)$ and the results will follow accordingly. Although the physical interpretations remain unaltered. 
        
        In case of exoplanetary atmosphere modeling, the atmospheric emission can not always be simplified by the Planck emission. The upper atmosphere of exoplanets does not hold the local thermodynamic equilibrium condition \citep{seager2010exoplanet,sengupta2021effects}. In such cases different types of atmospheric emission can be considered by the general atmospheric emission function $\beta$ as shown in \eqref{eq: general emission term}. By appropriate choice of $\beta$-parameter thermal re-emission, anisotropic emission can be considered as discussed in \cite{sengupta2021effects}.
        
     Finally, the polarization effect is not considered in this work. That can be included for finite atmosphere in the same way as in the semi-infinite atmosphere case discussed in \cite{sengupta2021effects}.


\appendix

\section{DERIVATION OF THE SCATTERING FUNCTION and its derivative}\label{apndx sec: Derivation of scattering function} 
In this section we will show the derivation of the scattering function $S(\tau_1,\mu,\phi;\mu_0,\phi_0)$ and its derivative $\frac{\partial S(\tau,\mu,\phi;\mu_0,\phi_0)}{\partial \tau}|_{\tau = \tau_1}$ which has directly been written in eqns.\eqref{eq: finite atmosphere scattering function1.1} and \eqref{eq: finite atmosphere scattering function2.1}.For the simplicity of calculations we write $\frac{\partial S(\tau_1,\mu,\phi;\mu_0,\phi_0)}{\partial \tau_1}$ instead of $\frac{\partial S(\tau,\mu,\phi;\mu_0,\phi_0)}{\partial \tau}|_{\tau = \tau_1}$ in the main text of section~\ref{sec: general integral equations}. To start with we will use the scattering function relation with the source function equation given
in \cite{chandrasekhar1960radiative} (page: 168; eqns: (23) and (25)) as follows,

\begin{equation}\label{eq: scattering function in chandrasekhar 23}
    \frac{1}{4}[(\frac{1}{\mu}+\frac{1}{\mu_0})S(\tau_1;\mu,\phi;\mu_0,\phi_0) + \frac{\partial S(\tau;\mu,\phi;\mu_0,\phi_0)}{\partial \tau}|_{\tau = \tau_1}] = \xi(0,+\mu,\phi) + \frac{1}{4\pi}\int_0^1\int_0^{2\pi} S(\tau_1;\mu,\phi;\mu',\phi')\xi(0,-\mu',\phi') \frac{d\mu'}{\mu'} d\phi'
\end{equation}

and 

\begin{equation}\label{eq: scattering function in chandrasekhar 25}
    \frac{1}{4}F \frac{\partial S(\tau;\mu,\phi;\mu_0,\phi_0)}{\partial \tau}|_{\tau = \tau_1} = exp(-\tau_1/\mu)\xi(\tau_1,+\mu,\phi) + \frac{1}{4\pi} \int_0^1\int_0^{2\pi} T(\tau_1;\mu,\phi;\mu',\phi')\xi(\tau_1,+\mu',\phi')\frac{d\mu'}{\mu'}d\phi'
\end{equation}

Now the source functions $\xi(0,\mu,\phi)$ and $\xi(\tau_1,\mu,\phi)$ in thermal emission case is given in eqns. \eqref{eq: source function at tau=0} and \eqref{eq: source function at tau=tau_1}. Making
use of them with the boundary conditions (eqns.\eqref{eq: boundary condition1}) the above two equations can be written as,

\begin{equation}\label{apndx eq: scattering function}
\begin{split}
    &\frac{F}{4}[(\frac{1}{\mu}+\frac{1}{\mu_0})S(\tau_1;\mu,\phi;\mu_0,\phi_0)+\frac{\partial S(\tau;\mu,\phi;\mu_0,\phi_0)}{\partial \tau}|_{\tau = \tau_1}]=
B(T_0)[1+\frac{1}{4\pi}\int_0^1\int_0^{2\pi} S(\tau_1;\mu,\phi;\mu',\phi')\frac{d\mu'}{\mu'}d\phi']\\
&+
\frac{1}{4}F[p(\mu,\phi;-\mu_0,\phi_0)+ \frac{1}{4\pi}\int_0^1\int_0^{2\pi} p(\mu,\phi;\mu'';\phi'')S(\tau_1;\mu'',\phi'';\mu_0,\phi_0)d\phi''\frac{d\mu''}{\mu''}\\
&+
\frac{1}{4\pi}\int_0^1\int_0^{2\pi} S(\tau_1;\mu,\phi;\mu',\phi')\{p(-\mu',\phi';-\mu_0,\phi_0)+
\frac{1}{4\pi}\int_0^1\int_0^{2\pi} p(-\mu',\phi';\mu'';\phi'')S(\tau_1;\mu'',\phi'';\mu_0,\phi_0)d\phi''\frac{d\mu''}{\mu''}\}\frac{d\mu'}{\mu'}d\phi']
\end{split}
\end{equation}

and 

\begin{equation}\label{apndx eq: derivative scattering function}
    \begin{split}
        &\frac{F}{4}[\frac{\partial S(\tau;\mu,\phi;\mu_0,\phi_0)}{\partial \tau}|_{\tau = \tau_1}]=
B(T_{\tau_1})[e^{-\tau_1/\mu}+\frac{1}{4\pi}\int_0^1\int_0^{2\pi} T(\tau_1;\mu,\phi;\mu',\phi')\frac{d\mu'}{\mu'}d\phi']\\
&+
\frac{1}{4}F[exp\{-\tau_1(\frac{1}{\mu_0}+\frac{1}{\mu})\}p(\mu,\phi;-\mu_0,\phi_0)
+
\frac{1}{4\pi}e^{-\tau_1/\mu}\int_0^1\int_0^{2\pi} p(\mu,\phi;-\mu'';\phi'')T(\tau_1;\mu'',\phi'';\mu_0,\phi_0)d\phi''\frac{d\mu''}{\mu''}\\
&+
\frac{1}{4\pi}\int_0^1\int_0^{2\pi} T(\tau_1;\mu,\phi;\mu',\phi')\{e^{-\tau_1/\mu_0}p(\mu',\phi';-\mu_0,\phi_0)
+
\frac{1}{4\pi}\int_0^1\int_0^{2\pi} p(\mu',\phi';-\mu'';\phi'')T(\tau_1;\mu'',\phi'';\mu_0,\phi_0)d\phi''\frac{d\mu''}{\mu''}\} \frac{d\mu'}{\mu'}d\phi']
    \end{split}
\end{equation}

Hence, multiplying eqns.\eqref{apndx eq: scattering function} and \eqref{apndx eq: derivative scattering function} by $\frac{4}{F}$ and replacing the quantities $\frac{B(T_0)}{F}, \frac{B(T_{\tau_1})}{F}$ by $U(T_0)$ and $U(T_{\tau_1})$ respectively, we will 
get the equations \eqref{eq: finite atmosphere scattering function1.1} and \eqref{eq: finite atmosphere scattering function2.1} respectively. In the similar fashion the Transmission function $T(\tau_1;\mu,\phi;\mu_0,\phi_0)$ and its derivative $\frac{\partial T(\tau;\mu,\phi;\mu_0,\phi_0)}{\partial \tau}|_{\tau = \tau_1}$ can be derived.


\bibliography{paper3}
\bibliographystyle{aasjournal}

\end{document}